\documentclass[aps,prb,groupedaddress]{revtex4}
\usepackage{graphics}
\begin{document}

\title{Two-dimensional 
negative donors in magnetic fields}

\author{Mikhail V. Ivanov}
\email{mivanov@mi1596.spb.edu}
\altaffiliation[Permanent address: ]{Institute of Precambrian Geology 
and Geochronology, Russian Academy of Sciences, Nab. Makarova 2, 
St. Petersburg 199034, Russia }

\author{Peter Schmelcher}
\email{Peter.Schmelcher@tc.pci.uni-heidelberg.de}
\affiliation{Theoretische Chemie, Physikalisch-Chemisches Institut, 
Universit\"at Heidelberg, INF 229, D-69120 Heidelberg, 
Federal Republic of Germany}

\date{\today}

\begin{abstract}
A finite-difference solution of the Schr\"odinger equation 
for negative donor centers $D^-$ in two dimensions is presented. 
Our approach is of exact nature and allows us to resolve a discrepancy
in the literature on the ground state of a negative donor.
Detailed calculations of the energies for a number of states  
show that for field strengths less than $\gamma=0.117 a.u.$ the donor possesses
one bound state, for $0.117<\gamma<1.68 a.u.$ there exist two bound states 
and for field strengths $\gamma \geq 1.68 a.u.$ the system possesses three bound states. 
Further relevant characteristics of negative donors in magnetic fields are
provided.
\end{abstract}


\maketitle

\section{Introduction\label{secintro}}

The properties of $D^-$ centers in narrow quantum wells have become a subject of
considerable interest during the past years \cite{Phelps,Larsen1,Larsen2,XuWang,Sandler,TaoPang,Louie,Zhu1,Zhu2}. 
In very narrow wells this system can be considered as a two-dimensional counterpart
of the atomic ${\rm H^-}$ ion. A frequently used compound for current experimental investigations 
(see for example ref.\cite{Jiang}) of such systems are layers of GaAs/AlGaAs. 
The high mobility i.e. small effective mass of the electrons and the comparatively large
dielectricity constant of this semiconductor material allow us to study strong magnetic
field effects in the laboratory. Both the two-dimensional character of the motion of the electrons 
and the external magnetic field makes the ground state of the system 
more tightly bound than that of the field-free threedimensional ${\rm H^-}$ ion. Considering
neutral donors $D^0$ the planar electron density of the 2d donor is enhanced significantly
compared to that of the 3d donor. 
The form of the corresponding charge distribution makes the interaction of an additional distant charge with the neutral donor very different for the three-dimensional compared to the two-dimensional situation. 
In three dimensions
it is well-known that, for a fixed center (nucleus),
the combination of a magnetic field with the 3d long-range attractive
polarization forces lead to an infinite number of bound states for the negative donor
\cite{Avr81,Bez00,Bez01,AlH00}. In two dimensions the long-range interaction
is of repulsive character. Indeed, it was shown by Larsen and McCann \cite{Larsen1}, that an electron situated 
far from a 2d $D^0$ center experiences an overall repulsive potential which prevails
both the attractive Coulomb attraction due to the center and the attractive polarization forces. 
For distances of the order of the extension of the neutral donor, exchange and correlation effects play a major role with respect to the binding mechanism of the additional electron. 

Two-dimensional $D^-$ centers in magnetic fields are described by a three-dimensional
Schr\"odinger equation. 
The finite difference numerical method 
\cite{Ivanov86,Ivanov85,Ivanov88,Ivanov94,Ivanov98,IvaSchm2001b,IvaSchm2001c,Ivanov86a}
allows us to solve this Schr\"odinger equation without any simplifications or approximations
with respect to the geometry of the wave function, correlation effects etc.
Due to this property of our computational approach it will be here possible
to obtain a definite answer on the ground state energy
and the number of bound states of the two-dimensional $D^-$ center in magnetic fields.
The critical values of the field strengths that are associated with the appearance of the bound states will be determined. 
One of the major motivations for the present work are the
different theoretical values for the ground state energy of a two-dimensional
negative donor with and without magnetic field existing in the literature.
Variational calculations \cite{Phelps,Larsen2} on the onehand and Monte Carlo simulations
\cite{Louie} on the otherhand yield uncompatible predictions for the ground state energies.
This puzzle will be resolved within the present work.

\section{The Schr\"odinger equation and the method of solution\label{seceq}}

The Hamiltonian of our two-dimensional system of two interacting electrons 
with an effective mass $m$ and a singly charged positive ion 
placed into a magnetic field perpendicular to the plane can be 
written in Cartesian coordinates $(x,y)$ as 
\begin{eqnarray}
H=-\frac 12 (\nabla^2_1+\nabla^2_2)
-i\frac\gamma 2\left(
-y_1\frac\partial{\partial x_1}+x_1\frac\partial{\partial y_1}
-y_2\frac\partial{\partial x_2}+x_2\frac\partial{\partial y_2}
\right) 
+\frac {\gamma^2}8(r_1^2+r_2^2)\nonumber\\
-\frac 1 {r_1}-\frac 1 {r_2}+\frac 1 {r_{12}}
\label{eq:Ham01}
\end{eqnarray}
Here we use the units 
$a_{\rm eff}=\hbar^2 \epsilon e^{-2} m^{-1}=\epsilon m^{*-1}\cdot 5.29\times 10^{-9}$cm
for the length,  
$E_{\rm eff}=e^4\hbar^{-2}m\epsilon^{-2}=m^*\epsilon^{-2}\cdot 27.2$eV 
for the energy and 
$B_{\rm eff}=ce^3m^2\hbar^{-3}\epsilon^{-2}=m^{*2}\epsilon^{-2}\cdot 
2.3505\times 10^5$T for the magnetic field strength.  
$\gamma=B/B_{\rm eff}$, $m_0$ is the mass of the free electron, 
$m^*=m/m_0$, $\epsilon$ is the dielectricity constant of the semiconductor material, 
$r_1=(x_1^2+y_1^2)^{1/2}$, $r_2=(x_2^2+y_2^2)^{1/2}$,
and $r_{12}=\left|{\vec r}_1 - {\vec r}_2  \right|$. 
The spin terms are omitted.

In contrast to variational calculations we are not biased by using a particular
ansatz for the variational wave function but will perform a full grid
solution to the Schr\"odinger equation that allows to control and estimate
the remaining minor deviation from the exact eigenfunctions and eigenvalues.
In the following we provide an outline of our approach which embodies various
properties and transformations of the Hamiltonian and the corresponding Schr\"odinger
equation.

Transforming the electronic degrees of freedom to polar coordinates $(r_1,\phi_1)$ and $(r_2,\phi_2)$ 
yields for the Hamiltonian (\ref{eq:Ham01}) 

\begin{eqnarray}
H=-\frac 12\left(
\frac{\partial ^2}{\partial r_1 ^2}+
\frac 1 {r_1} \frac \partial{\partial r_1 }
+\frac{\partial ^2}{\partial r_2 ^2}+
\frac 1 {r_2} \frac \partial{\partial r_2 }
\right)
-\frac 12\left(
\frac 1 {r_1^2} \frac {\partial^2}{\partial \phi_1^2 }+
\frac 1 {r_2^2} \frac {\partial^2}{\partial \phi_2^2 }
\right)
-i\frac \gamma 2 \left(
\frac {\partial}{\partial \phi_1 }
+\frac {\partial}{\partial \phi_2 }
\right)
\nonumber\\
+\frac{\gamma^2}{8}(r_1^2+r_2^2)
-\frac 1{r_1}-\frac 1{r_2}
+\frac 1 {(r_1^2+r_2^2-2r_1r_2\cos (\phi_1 - \phi_2))^{1/2}}
\label{eq:Ham02}
\end{eqnarray}

The eigenfunctions of this Hamiltonian are the eigenfunctions of the 
Hamiltonian (\ref{eq:Ham01}) if we require them to be periodic with period $2\pi$ 
with respect to the variables $\phi_1$ and $\phi_2$. 
The eigenfunctions of the Hamiltonian (\ref{eq:Ham02}) 
should also be eigenfunctions of the $z$-projection of the total orbital angular momentum operator
which is a conserved quantity ie. ${\bf l}_z\Psi=M\Psi\nonumber$ with
\begin{eqnarray}
{\bf l}_z=
-i\frac{\partial }{\partial \phi_1 }-i\frac{\partial }{\partial \phi_2 }
\label{eq:Eq06}
\end{eqnarray}
Introducing the coordinates $r_1,r_2,\phi,\Phi$ where
\begin{eqnarray}
\phi=\phi_1-\phi_2
\label{eq:Eq008}
\end{eqnarray}
and
\begin{eqnarray}
\Phi=\frac{\phi_1+\phi_2}{2}
\label{eq:Eq009}
\end{eqnarray}
we have
\begin{eqnarray}
{\bf l}_z=-i\frac{\partial }{\partial \Phi }
\label{eq:Eq007}
\end{eqnarray}
For these coordinates the Hamiltonian takes on the appearance
\begin{eqnarray}
H=-\frac 12\left(
\frac{\partial ^2}{\partial r_1 ^2}+
\frac 1 {r_1} \frac \partial{\partial r_1 }
+\frac{\partial ^2}{\partial r_2 ^2}+
\frac 1 {r_2} \frac \partial{\partial r_2 }
\right)
-\frac 18\left(\frac 1 {r_1^2} +\frac 1 {r_2^2}\right)
\frac{\partial^2}{\partial\Phi^2}
\nonumber\\
-\frac 12\left(\frac 1{r_1^2} -\frac 1{r_2^2}\right)
\frac{\partial^2}{\partial\Phi\partial\phi}
-\frac 12\left(\frac 1 {r_1^2} +\frac 1 {r_2^2}\right)
\frac{\partial^2}{\partial\phi^2}
-i\frac \gamma 2 \frac {\partial}{\partial \Phi }
\nonumber\\
+\frac{\gamma^2}{8}(r_1^2+r_2^2)
-\frac 1{r_1}-\frac 1{r_2}
+\frac 1 {(r_1^2+r_2^2-2r_1r_2\cos \phi)^{1/2}}
\label{eq:Ham10}
\end{eqnarray}
Eigenfunctions of the operator ${\bf l}_z$ read as follows
\begin{eqnarray}
\Psi(r_1,r_2,\phi,\Phi)=e^{iM\Phi}\psi(r_1,r_2,\phi)
\label{eq:Phiphi}
\end{eqnarray}
We therefore have
\begin{eqnarray}
H\Psi=e^{iM\Phi}h\psi
\end{eqnarray}
with
\begin{eqnarray}
h=-\frac 12\left(
\frac{\partial ^2}{\partial r_1 ^2}+
\frac 1 {r_1} \frac \partial{\partial r_1 }
+\frac{\partial ^2}{\partial r_2 ^2}+
\frac 1 {r_2} \frac \partial{\partial r_2 }
\right)
+\frac {M^2}8\left(\frac 1 {r_1^2}+\frac 1 {r_2^2}\right)
\nonumber\\
-\frac{iM}2\left(\frac 1 {r_1^2}-\frac 1 {r_2^2}\right)
\frac {\partial}{\partial \phi }
-\frac 12\left(\frac 1 {r_1^2}+\frac 1 {r_2^2}\right)
\frac {\partial^2}{\partial \phi^2 }
+\frac {M\gamma} 2 
\nonumber\\
+\frac{\gamma^2}{8}(r_1^2+r_2^2)
-\frac 1{r_1}-\frac 1{r_2}
+\frac 1 {(r_1^2+r_2^2-2r_1r_2\cos \phi)^{1/2}}
\label{eq:Ham06}
\end{eqnarray}
This Hamiltonian depends on three degrees of freedom $(r_1,r_2,\phi)$
and the corresponding Schr\"odinger equation can therefore be solved for
the eigenfunctions and eigenvalues by applying 
our finite-difference method \cite{Ivanov86, Ivanov85,Ivanov88}.
The coordinates $r_1$ and $r_2$ are in the domain
\begin{eqnarray}
0<r_1<+\infty,\ \ 0<r_2<+\infty
\end{eqnarray}
with zero boundary conditions for infinity. The corresponding domain and boundary conditions for
the angle $\phi$ have to be analyzed in more detail.

From equation (\ref{eq:Eq009}) one can conclude 
that solutions of the Schr\"odinger equation belonging to
the Hamiltonian (\ref{eq:Ham06}) should be considered 
in the domain $-2\pi\leq\phi\leq2\pi$ with cyclic boundary
conditions. This leads to eigenfunctions that can be either symmetric
or antisymmetric with respect to interparticle exchange.
The first correspond to spin singlet states whereas the second yield spin triplet states. 
Beyond this we encounter additional eigenfunctions of the operator (\ref{eq:Ham06})
that are not eigenfunctions of our initial Hamiltonian (\ref{eq:Ham01}) i.e.
these eigenfunctions do not describe physical solutions.
Indeed, exploiting certain symmetries of the wavefunction it is possible
to confine the domain of the angle $\phi$ to $0\leq\phi\leq2\pi$ for
the Schr\"odinger equation belonging to the Hamiltonian (\ref{eq:Ham06}).
As a consequence the non-physical solutions are excluded. 
For the states with $M=0$ solutions can be obtained in an even smaller domain $0\leq\phi\leq\pi$. 

The numerical finite-difference method employed here to solve the
eigenvalue problem for the Hamiltonian (\ref{eq:Ham06})
is a modification of the approach developed in previous works devoted to atoms in strong 
magnetic \cite{Ivanov88,Ivanov94,IvaSchm2001b} and electric fields 
(see\cite{Ivanov98,IvaSchm2001c} and references therein) 
and contains, in particular, technical aspects used in three-dimensional problems \cite{Ivanov85,Ivanov86a}. 
Our computational procedure consists of the following main steps. 
The nodes have to be chosen in the spatial domain 
$\Omega:0\leq r_1<+\infty$, $0\leq r_2<+\infty$, $0\leq\phi\leq 2\pi$ (or $0\leq\phi\leq\pi$).
The values of the wavefunctions at the positions of the nodes are the numerical 
representation of the solutions of the initial differential equations. 
The nodes of the three-dimensional mesh in the space $(r_1,r_2,\phi)$ 
are placed at all the points with coordinates $(r_{1i},r_{2j},\phi_{k})$ where 
$r_{1i}$, $r_{2j}$, and $\phi_{k}$ are sets of the mesh node coordinates 
along the corresponding axes. For the coordinate $\phi$ it is natural
to use a uniform mesh with nodes at 
$\phi_{k}=2\pi(k-1/2)/N$ for the region $0\leq\phi\leq2\pi$ (case $M \neq 0$) and 
$\phi_{k}=\pi(k-1/2)/N$ for the region $0\leq\phi\leq\pi$ (case $M=0$), 
where $N$ is the number of nodes in the direction $\phi$. 
For $r_1$ and $r_2$ we have used non-uniform distributions of nodes 
similar to those described in ref.\cite{IvaSchm2001b}, 
which cover the infinite domains $0\leq r_1<+\infty$, $0\leq r_2<+\infty$ 
as $N\rightarrow\infty$. 

We employ the inverse iteration method to obtain the eigenfunctions and energy eigenvalues.
This requires solving a system of linear equations 
with a matrix that is a finite-difference approximation to the Hamiltonian. 
The solution of these equations is particularly simple if the matrix has a block-tridiagonal structure. 
The latter can be achieved using the simplest three-point approximation 
for the derivatives for one of the coordinates. 
The choice of this coordinate is dictated by obtaining a convenient form of representation for the boundary conditions. 
For the two other coordinates we are free to apply more precise higher order approximations 
to their derivatives. The final values for the energy (and other observables) are provided by
using the Richardson extrapolation technique for the corresponding results emerging from a series 
of geometrically similar meshes with different number of nodes. 
Using this approach we achieve a major increase of the numerical precision 
and, in particular, we obtain together with each numerical value a reliable estimate 
of its precision\cite{Ivanov86,Ivanov88}. 
Typically, meshes used in the present calculations range from
(the sparsest) mesh with $20^3$ nodes to (the thickest) one with $38^3$ 
nodes, i.e. 38 nodes in each direction.

\section{Results and discussion}

Our results for the energies of the ground state of the two-dimensional $D^-$ center 
are presented in table \ref{tab:grcon}.
This table contains also the corresponding energies for the ground state of 
the $D^0$ center and a comparison with the best results 
obtained in the literature \cite{Larsen1,Larsen2,Louie}. Ref.\cite{Larsen1,Larsen2} are
variational and ref.\cite{Louie} represent Monte-Carlo simulations.
One can see, that the variational results by Larsen and McCann coincide for weak and intermediate field
strengths \cite{Larsen2} very well with our results. In contrast to this
the results obtained by Louie and Tao Pang\cite{Louie} differ significantly from these values.
The absence of any approximations in our approach and
the possibility to reliably evaluate the convergence of our results allow us to 
conclude, that the results obtained in ref.\cite{Louie} overestimate the true values for
the binding energy of the ground state of the two-dimensional negative donor considerably. 
(The reader should note that all digits of the values for our calculated energies
given in table \ref{tab:grcon} are converged i.e. coincide with the exact results). This resolves the discrepancy
on the ground state energy of the negative donor present in the literature as
demonstrated by the results contained in table \ref{tab:grcon} particularly for $\gamma =0$
but also for nonvanishing field strengths.

\begin{table}
\caption{$D^-$ donor total and binding energies in two dimensions for the $M=0$ 
singlet ground state. Results of our calculations [IS] and 
refs.\cite{Larsen2,Larsen1,Louie} [other]. The energies of the neutral donor are also given.}
\begin{ruledtabular}
\begin{tabular}{lllllll}
$\gamma$&$E^{D^-}$[IS]&$E^{D^-}$[other]&$E^D$[IS]&$E^D$[other]&
$E^{D^-}_B$[IS]&$E^{D^-}_B$[other]\\ \noalign{\hrule}
0     &$-2.24027$&$-2.239$\cite{Larsen2}&$-2.00000$&$-
2.000$\cite{Larsen2}&0.24027&$0.239$\cite{Larsen2}\\
      &          &&&$-2.000$\cite{Louie}&&$0.2555$\cite{Louie}\\
0.02  &$-2.26014$&&$-2.009967 $&&0.25017\\
0.05  &$-2.28947$&&$-2.024869 $&&0.26460\\
0.1   &$-2.33712$&&$-2.049518 $&&0.28760\\
0.2   &$-2.42808$&&$-2.098116 $&&0.32996\\
0.5   &$-2.67354$&$-2.673$\cite{Larsen2}&$-2.238416 $&$-
2.239$\cite{Larsen2}&0.43512&$0.435$\cite{Larsen2}\\
1.0   &$-3.02151$&$-3.021$\cite{Larsen2}&$-2.455152 $&$-
2.455$\cite{Larsen2}&0.56636&$0.566$\cite{Larsen2}\\
1.0   &&&$        $&$-2.455$\cite{Louie}&&$0.585$\cite{Louie}\\
2.0   &$-3.58733$&$-3.586$\cite{Larsen2}&$-2.836203 $&$-
2.836$\cite{Larsen2}&0.75113&$0.750$\cite{Larsen2}\\
3.0   &$-4.05355$&&$-3.165976$&$-
3.175$\cite{Louie}&$0.88757$&$0.91$\cite{Louie}\\
4.0   &$-4.45883$&$-4.459$\cite{Larsen2}&$-3.459582 $&$-
3.459$\cite{Larsen2}&0.99925&$1.000$\cite{Larsen2}\\
10.   &$-6.27690$&$-6.261$\cite{Larsen2}&$-4.815151 $&$-
4.813$\cite{Larsen2}&1.46175&$1.463$\cite{Larsen2}\\
20.   &$-8.36994$&$-8.369$\cite{Larsen2}&$-6.407114 $&$-
6.405$\cite{Larsen2}&1.96283&$1.964$\cite{Larsen2}\\
50.&$-12.5581$&$-11.4583$\cite{Larsen1}&$-9.62189$&$-
8.86227$\cite{Larsen1}&2.9362&$2.5961$\cite{Larsen1}\\ 
100.  &$-17.2937$&$-16.2045$\cite{Larsen1}&$-13.27207 $&$-
12.53314$\cite{Larsen1}&4.0216&$3.6715$\cite{Larsen1}\\
200.  &$-23.9989$&$-22.9166$\cite{Larsen1}&$-18.449164$&$-
17.72454$\cite{Larsen1}&5.5497&$5.1923$\cite{Larsen1}\\
500.  &$-37.3107$&$-36.234$\cite{Larsen1}&$-28.737120$&$-
28.02495$\cite{Larsen1}&8.5736&$8.2097$\cite{Larsen1}\\
1000. &$-52.316 $&$-51.243$\cite{Larsen1}&$-40.33926 $&$-
39.63327$\cite{Larsen1}&11.977&$11.610$\cite{Larsen1}\\
\end{tabular}
\end{ruledtabular}
\label{tab:grcon}
\end{table}

For strong magnetic fields $\gamma>>1$ table~\ref{tab:grcon} provides also the energies
obtained by Larsen and McCann in another variational calculation\cite{Larsen1}. 
Our results both on the total and binding energies show
that the ground state of the system is more strongly bound than predicted by ref.\cite{Larsen1}. 
The differences for the energies depend only weakly on the magnetic field strength 
and are approximately 1.08~a.u. for the total energy and approximately 0.36~a.u. for the binding energy!
The significant difference in deviations of our results from those obtained in ref.\cite{Larsen1} is due to the fact that the binding energies of the neutral donor in a strong magnetic field are underestimated in ref.\cite{Larsen1}. 
As shown below this leads also to an overestimation of the binding properties 
of the excited states of the $D^-$ center.

\begin{table}
\caption{Total and binding energies for 
$D^-$ triplet excited states with $M<0$.}
\begin{ruledtabular}
\begin{tabular}{lllllllllllll}
$\gamma$&
$E^{D^-}_{M=-1}$&$E_B$&
$E^{D^-}_{M=-2}$&$E_B$&
$E^{D^-}_{M=-3}$&$E_B$&\\
\noalign{\hrule}
0.05 &$-2.02453 $&$-0.00034$&$-2.02474$&$-0.00013$&$-2.02481$&$-0.00006$\\
0.1  &$-2.0492  $&$-0.0003 $&$-2.04916$&$-0.00036$&$-2.04933$&$-0.00019$\\
0.2  &$-2.100   $&$0.002   $&$-2.09715$&$-0.00097$&$-2.09759$&$-0.000528 $\\
0.5  &$-2.253   $&$0.015   $&$-2.2360 $&$-0.0024 $&$-2.236554$&$-0.001862$\\
1.0  &$-2.50275 $&$0.04760 $&$-2.4523 $&$-0.0029 $&$-2.4511 $&$-0.0039   $\\
2.0  &$-2.9512  $&$0.1150  $&$-2.83775$&$0.00155 $&$-2.8302 $&$-0.0060   $\\
4.0  &$-3.68474 $&$0.2249  $&$-3.472622$&$0.01304$&$-3.450805$&$-0.008773$\\
10.  &$-5.2634  $&$0.4482  $&$-4.85787$&$0.04272 $&$-4.796  $&$-0.019    $\\
20.  &$-7.1028  $&$0.6957  $&$-6.4853 $&$0.0782  $&$-6.381  $&$-0.026    $\\
50.  &$-10.80161$&$1.17972 $&$-9.7707 $&$0.1488  $&$-9.590  $&$-0.032    $\\
100. &$-14.99309$&$1.72102 $&$-13.501 $&$0.229   $&$-13.236 $&$-0.036    $\\
200. &$-20.933  $&$2.484   $&$-18.794 $&$0.345   $&$-18.41  $&$-0.04     $\\
500. &$-32.732  $&$3.995   $&$-29.25  $&$0.51    $&$-28.69  $&$-0.05     $\\
500. &$-32.1289$\cite{Larsen1}&$4.1009$\cite{Larsen1}&$-
28.6261$\cite{Larsen1}&$0.602$\cite{Larsen1}&$-
28.0464$\cite{Larsen1}&$+0.021$\cite{Larsen1}\\
\end{tabular}
\end{ruledtabular}
\label{tab:grcon2}
\end{table}

The ground state of the $D^-$ center is the only bound state for $\gamma=0$. 
We have carried out calculations for excited states of both symmetric 
and antisymmetric character of the spatial wavefunction with respect to the interchange 
of the coordinates of the electrons (i.e. spin singlet and triplet states). 
These calculations were implemented both for $M=0$ and for $M\neq0$. 
The latter states were investigated also by Larsen and McCann\cite{Larsen1}. 
Without loss of generality we confine ourselves to negative magnetic quantum numbers $M<0$.
Pairs of states being different only with respect to the sign of $M$ 
possess equal binding energies within the corresponding Landau zones.
However, opposite to states with negative $M$, states with $M>0$ do 
not belong to the lowest Landau zone. Therefore, it is reasonable to
focus on values $M\leq0$. 

Our calculations show that the ground state of the $D^-$ center is the only spin 
singlet state which is bound in the presence of magnetic fields. 
This conclusion coincides with results obtained in ref.\cite{Larsen1}. 
On the otherhand, it follows from our calculations of the triplet states (see table 
\ref{tab:grcon2}) that there are two triplet states that become bound above some
corresponding critical values for $\gamma$. These are the energetically lowest states
for $M=-1$ and $M=-2$, respectively. 
The $M=-1$ state becomes bound for $\gamma>0.117$ (the total energy of both $D$ and 
$D^-$ at $\gamma=0.117$ is $-2.057852$). 
The state with $M=-2$ becomes bound for $\gamma>1.68$ (the total energy
at $\gamma=1.68$ is $-2.720788$). 

From table \ref{tab:grcon2} it is evident that the lowest state with magnetic
quantum number $M=-3$ is not bound for any magnetic field strength
considered here and probably also for higher field strengths (the binding energy $E_B=E^D-E^{D^-}$ is 
negative and its absolute value increases with increasing field strength). 
This result differs from that obtained by Larsen and McCann\cite{Larsen1}.
They received a positive binding energies for this state for sufficiently strong
magnetic fields. Their binding energy for $\gamma=500$ is presented
in the last row of table \ref{tab:grcon2}. 
The main reason for this discrepancy is the underestimation of the binding energy of 
the neutral donor in ref.\cite{Larsen1} as can be seen from table \ref{tab:grcon}. 
The data of table \ref{tab:grcon2} together with our results for other excited 
states (they are unbound) allow us to conclude that the $D^-$ 
considered possesses only two bound excited states -- possessing magnetic quantum
numbers $M=-1$ and $M=-2$.

\begin{figure}
\includegraphics{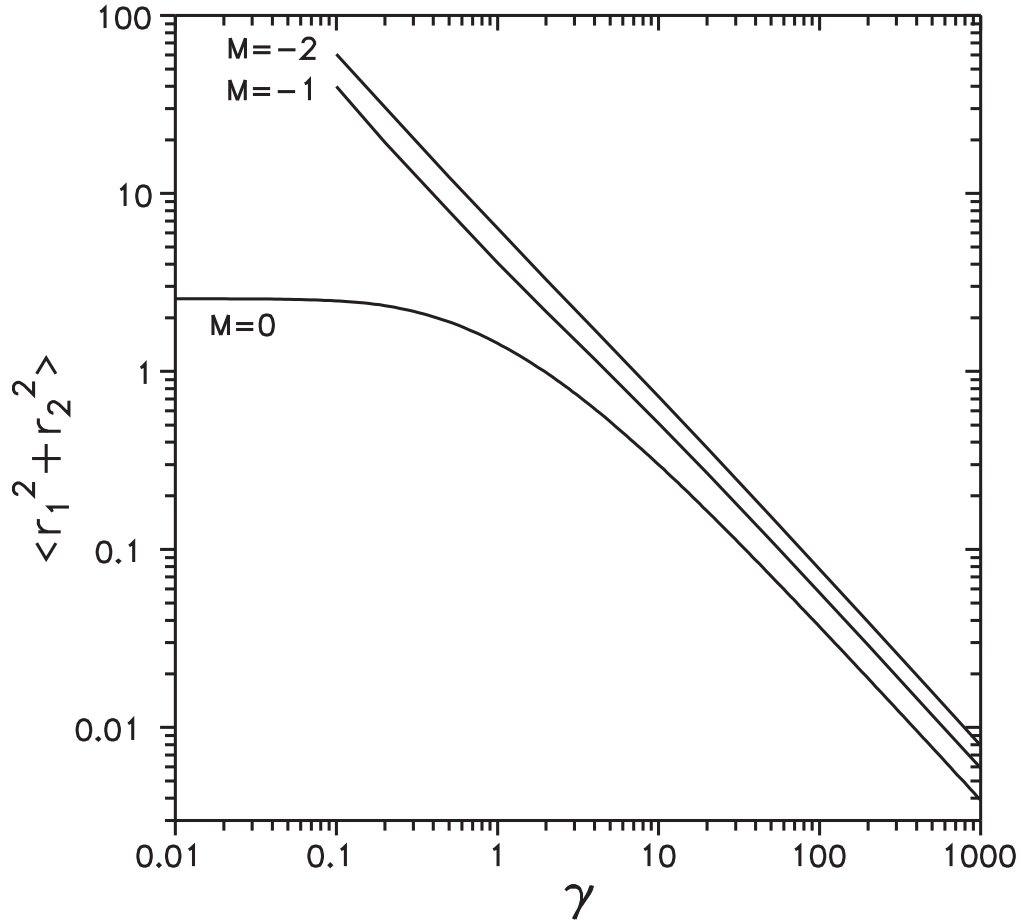}%
\caption{$\left<r_1^2+r_2^2\right>$ depending on the magnetic field strength for 
the three lowest states of the two-dimensional $D^-$ center. Effective atomic units
are used (see text).
\label{fig1}}
\end{figure}

Along with the total and binding energies of the two-dimensional negative donor 
$D^-$ we have calculated some geometrical parameters of its wavefunction that
provide additional information about the system. 
In figure~\ref{fig1} we present the expectation value $R^2=\left<r_1^2+r_2^2\right>$ 
which characterize the extension of the spatial distribution of the electrons in the simplest and 
most straightforward way as a function of the field strength. 
First of all we observe that the electronic cloud is shrinking monotonically 
with increasing field strength for all bound states. This had to expected
according to what we know about the behaviour of tightly bound states
of few-electron system in strong magnetic fields \cite{IvaSchm2001c}.
The difference of the behavior of $R^2$ for the ground ($M=0$) 
and excited ($M\neq0$) states for relatively weak magnetic fields is obvious:
The ground state is bound for $\gamma=0$, its wave function remains localized for 
all values of $\gamma$, and $R^2$ possesses a finite limit for
$\gamma\rightarrow0$. Furthermore it changes little for $\gamma \le 0.1a.u.$
For the excited states $R^2$ is not bounded for $\gamma\rightarrow0$. It 
changes rapidly with increasing field strength in particular in the weak
magnetic field regime. Therefore $R^2$ possesses finite values for excited states
only due to the presence of the magnetic field. 
On the otherhand, the dependencies of $R^2$ on the 
magnetic field strength in the high field regime are similar for all considered
states since they are dominated by the diamagnetic term $\gamma(r_1^2+r_2^2)/8$ of
the Hamiltonian. The occurence of a small curvature in the dependence 
$R^2(\gamma)$ for $M=-1$ between $\gamma=0.1$ and 
$\gamma=10$ is due to the influence of internal binding forces of the system, 
which are not negligibly small compared to the magnetic forces for this range of 
field strengths.

\section{Brief Summary}

We have shown that the two-dimensional negative donor $D^-$ possesses three
bound states ie. two bound excited states in the presence of a sufficiently
strong magnetic field. The spin singlet ground state is bound for arbitrary
field strengths. Our investigation of the ground state clarifies a discrepancy 
in the literature and confirms previous variational calculations
\cite{Phelps,Larsen2}. As a result the energy eigenvalues obtained in 
ref.\cite{Louie} turn out to violate the variational principle and are too low
i.e. the corresponding binding energies are too large. For weak and intermediate field strength
we obtain good agreement of our total and binding energies compared to those
of ref.\cite{Larsen2}. In the high field regime, however, a significant lowering
of the total energies and raising of the binding energies are obtained within
the present investigation. Our particular computational approach allows for
an estimate of the difference of our (up to several digits converged) results and the
exact ones which consequently
allows us to draw definite conclusions on the energies and properties of the donor.
A series of calculations for excited states show that two other states become 
bound with increasing magnetic field strength. 
They are the lowest excited states with $M=-1$ ($\gamma>0.117$) and $M=-2$ 
($\gamma>1.68$). The extension of these states decreases monotonically with
increasing field strength.

\begin{acknowledgments}
One of the authors (M.V.I.) gratefully acknowledges financial support by the 
Deutsche Forschungsgemeinschaft. This work was completed during a visit of P.S. 
to the University of Regensburg whose kind hospitality is appreciated. 
Financial support by the Graduiertenkolleg 'Nonlinearity and Nonequilibrium in Condensed Matter'
at the University of Regensburg is gratefully acknowledged.
\end{acknowledgments}

{}

\end{document}